%% file: paper.tex
\documentclass[review,number,sort&compress]{elsarticle}

\usepackage{lineno}

\linenumbers

\usepackage{amssymb}
\usepackage{ifthen} 
\usepackage{graphicx}
\usepackage{xspace}
\usepackage{amsfonts}
\usepackage{amsmath}
\usepackage{wasysym}
\usepackage{multirow}
\usepackage{psfrag}
\usepackage{url}
\usepackage{color}

\input{our-symbols-def}

\journal{Nuclear Instruments and Methods}

\begin{document}

\begin{frontmatter}



\title{Mass distributions marginalized over per-event errors}

\author[a]{D. Mart{\'{\i}}nez Santos}
\author[b]{F. Dupertuis}

\address[a]{NIKHEF and VU University Amsterdam, Amsterdam, The Netherlands}
\address[b]{Ecole Polytechnique F\'ed\'erale de Lausanne (EPFL), Lausanne, Switzerland}



\begin{abstract}
We present generalizations of the Crystal Ball function to describe mass peaks in which the per-event mass resolution is unknown and marginalized over.
The presented probability density functions are tested using a series of toy MC samples generated with Pythia and smeared with different amounts of multiple scattering and for different detector resolutions. 

\end{abstract}

\begin{keyword}
statistics \sep invariant mass peaks

\end{keyword}

\end{frontmatter}


\input{introduction}
\input{samples}
\input{apollonios}

\input{ghyp}
\input{offset}

\input{ranges}

\input{constraint}
\input{conclusions}
\input{ack}





\bibliographystyle{elsarticle-num}
\bibliography{paper}







\end{document}

%% file: our-symbols-def.tex
\newcommand{\Jpsimm}{\ensuremath{\Jpsi\to\mu^+\mu^-}\xspace}

\newcommand{\Jpsi}{\ensuremath{J/\psi}\xspace}

\newcommand{\BsJpsiPhi}{\ensuremath{B^0_s\to J/\psi \phi}\xspace}

\newcommand{\MeVcc}{\ensuremath{\,{\rm MeV}/c^2}\xspace}
\newcommand{\gevc}{\ensuremath{\,{\rm GeV}/c}\xspace}

\newcommand{\mevcc}{\ensuremath{\,{\rm MeV}/c^2}\xspace}

\newcommand{\pdf}{\ensuremath{p.d.f.}\xspace}

\newcommand{\figref}[1]{Fig.~\ref{#1}}
\newcommand{\tabref}[1]{Table~\ref{#1}}

\newcommand{\secref}[1]{Sect.~\ref{#1}}
\newcommand{\Ypsilon}{\Upsilon}
\newcommand{\RooFit}{{\bf RooFit}\xspace}

%% file: introduction.tex
\section{Introduction}
\label{sec:introduction}

A very common probability density function (\pdf) used to fit the mass peak of a resonance in experimental particle physics is the so-called Crystal Ball (CB) function~\cite{Oreglia:1980cs,Gaiser:1982yw,Skwarnicki:1986xj}:
\begin{equation}
\label{eq:CB}
p(m) \propto
\begin{cases}
e^{-\frac{1}{2} \left(\frac {m-\mu}{\sigma}\right)^2 } &\text{, if }  \frac{m-\mu}{\sigma} > -a\\
A\left(B-\frac{m-\mu}{\sigma}\right)^n &\text{, otherwise}
\end{cases}
\end{equation} 
\noindent where $m$ is the free variable (the measured mass), $\mu$ is the most probable value (the resonance mass), $\sigma$ the resolution, $a$ is called the
transition point and $n$ the power-law exponent. $A$ and $B$ are calculated by imposing the continuity of the function and its derivative at 
the transition point $a$.
This function consists of a Gaussian core, that models the detector resolution, with a tail on the left-hand side that parametrizes the effect
of photon radiation by the final state particles in the decay. In data analysis, one may deal with events which have different uncertainties on the measured mass, 
therefore distorting the core of the Crystal Ball, which will not be a Gaussian any more. 
This is sometimes modelled by the sum of two or three Crystal Ball functions, which is the equivalent of assuming that the per-event uncertainty is a sum of two or three delta functions.
However, per-event uncertainties are usually continuous functions very different from a sum of a small number of deltas.
One way of dealing with per-event uncertainties that follow a certain distribution, is to either make a \pdf conditional on the per-event uncertainty 
(if its distribution is known) or to perform the analysis in bins of the quantities that affect the per-event uncertainties (for example, particle momenta)
and combine them afterwards.
However, those procedures can significantly complicate the analysis, and in some cases one may prefer
to simply marginalize over the mass error and have a \pdf that describes the final mass peak, as:

\begin{equation}
p(m) \propto \int_0^{\infty} \frac{1}{\sqrt{v}}e^{-\frac{1}{2v}(m-\mu)^2}\rho(v)dv
\end{equation}

\noindent where $v$ is the variance and $\rho(v)$ the prior density of the variance.

In this paper, we will define some extensions of the Crystal Ball distribution for different assumptions on $\rho(v)$.
We will fit the proposed mass models to \Jpsimm toy MC samples where we can modify the relative importance 
of multiple scattering (MS) and detector spatial resolution (hereafter SR).
Section \ref{sec:samples} describes the generation of the toy MC samples. Section \ref{sec:apollonios} defines an
extension of the CB using a hyperbolic distribution core. Sections \ref{sec:ghyp} and \ref{sec:offset} generalize the function
defined in \secref{sec:apollonios}. Section \ref{sec:ranges} gives a brief discussion of the meaning of the fit parameters.
Section \ref{sec:const} discusses other effects on the invariant mass line-shape that are not directly related to resolution. 
Conclusions are drawn in \secref{sec:conclusions}.

%% file: samples.tex
\section{Simulation of \Jpsimm decays}
\label{sec:samples}

We generate \Jpsi events at $\sqrt{s}=$ 8 TeV using the \texttt{main17.cc} script of \texttt{Pythia8.176} \cite{pythia}. The \Jpsi's are then
isotropically decayed into two muons. No photons are added, as the radiative tail of the mass distribution
should be well accounted by the Crystal Ball tail.

The generated muon momenta are smeared with a Gaussian resolution which has a momentum dependence:
\begin{equation}
\label{eq:smearing}
\frac{\sigma(p)}{p} = a + b p
\end{equation}  

\noindent where $a$ mimics the multiple scattering (MS) and $b$ mimics the effect of the
hit resolution. We take as typical values $a = 3\times10^{-3}$ and $b = 2\times10^{-5} \ \mathrm{GeV}^{-1}c$ inspired by~\cite{Alves:2008zz}, although we will vary them for different tests.

%% file: apollonios.tex
\section{Hyperbolic resolution model}
\label{sec:apollonios}

\begin{figure}
\label{fig:amoroso}
\begin{center}
\includegraphics[width=0.8\textwidth]{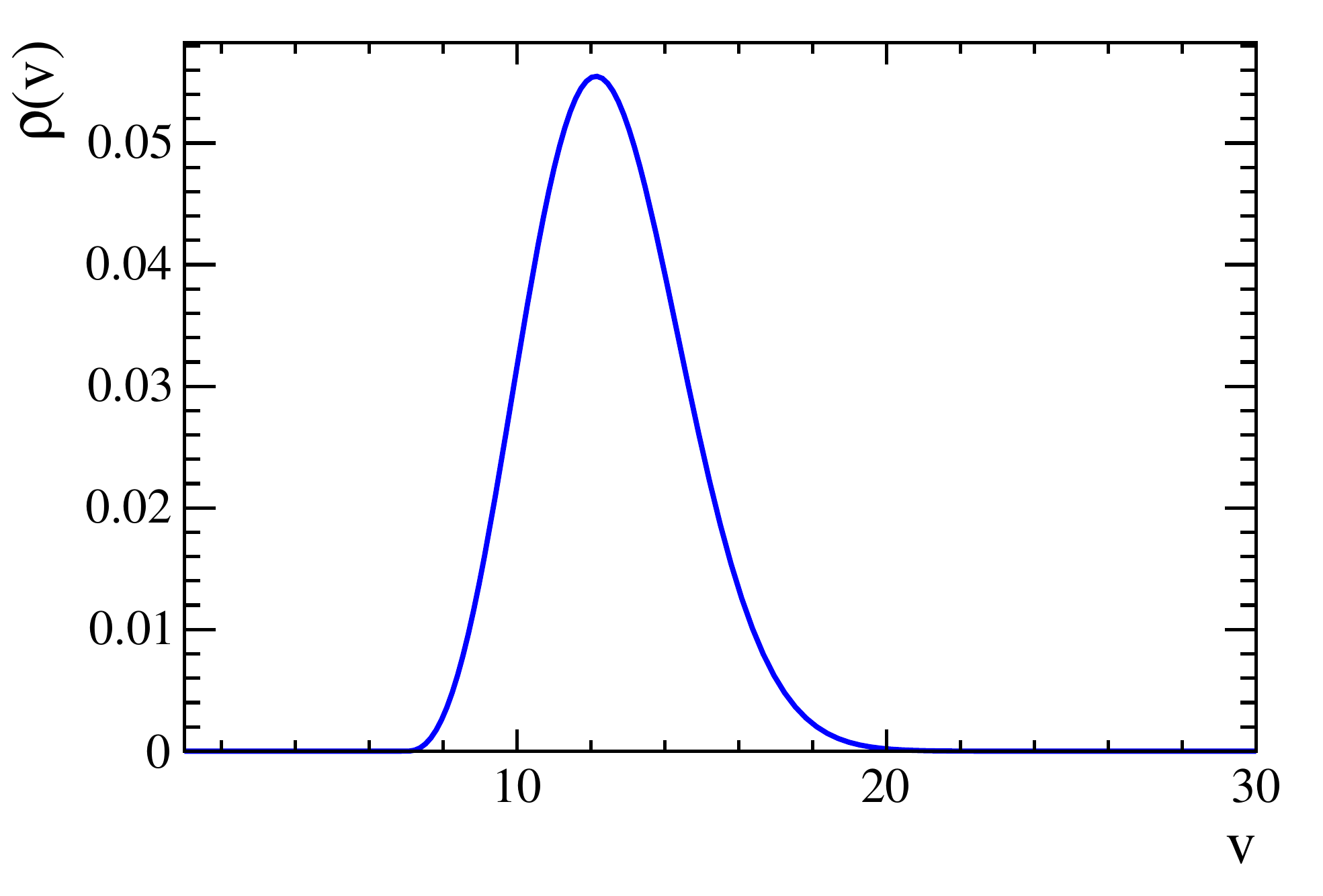}
\caption{Example of the Amoroso distribution for a hypothetical variance $v$. The parameters used are $\theta = 5$, $\alpha = 1.5$ $\mu = 7$, and $\beta = 2.3$. }
\end{center}
\end{figure}

A very flexible function that describes asymmetric unimodal \pdf's defined above a certain threshold (i.e, like per-event error distributions
usually look like, see for example Fig. 6 in \cite{Aaij:2013oba}) is the so-called Amoroso distribution~\cite{amoroso} (see \figref{fig:amoroso}). If we consider the Amoroso distribution as a potential
implementation for $\rho(v)$, then the corresponding core of the invariant mass \pdf will be the following:

\begin{equation}
\label{eq:phi}
\Phi(m) \propto  \int_0^{\infty} \frac{1}{\sqrt{v}}e^{-\frac{1}{2v}(m-\mu)^2}\left(\frac{v-v_0}{\theta}\right)^{\alpha\beta-1}e^{-\left(\frac{v-v_0}{\theta}\right)^{\beta}}dv
\end{equation}

\begin{figure}
\label{fig:fit1}
\begin{center}
\includegraphics[width=0.49\textwidth]{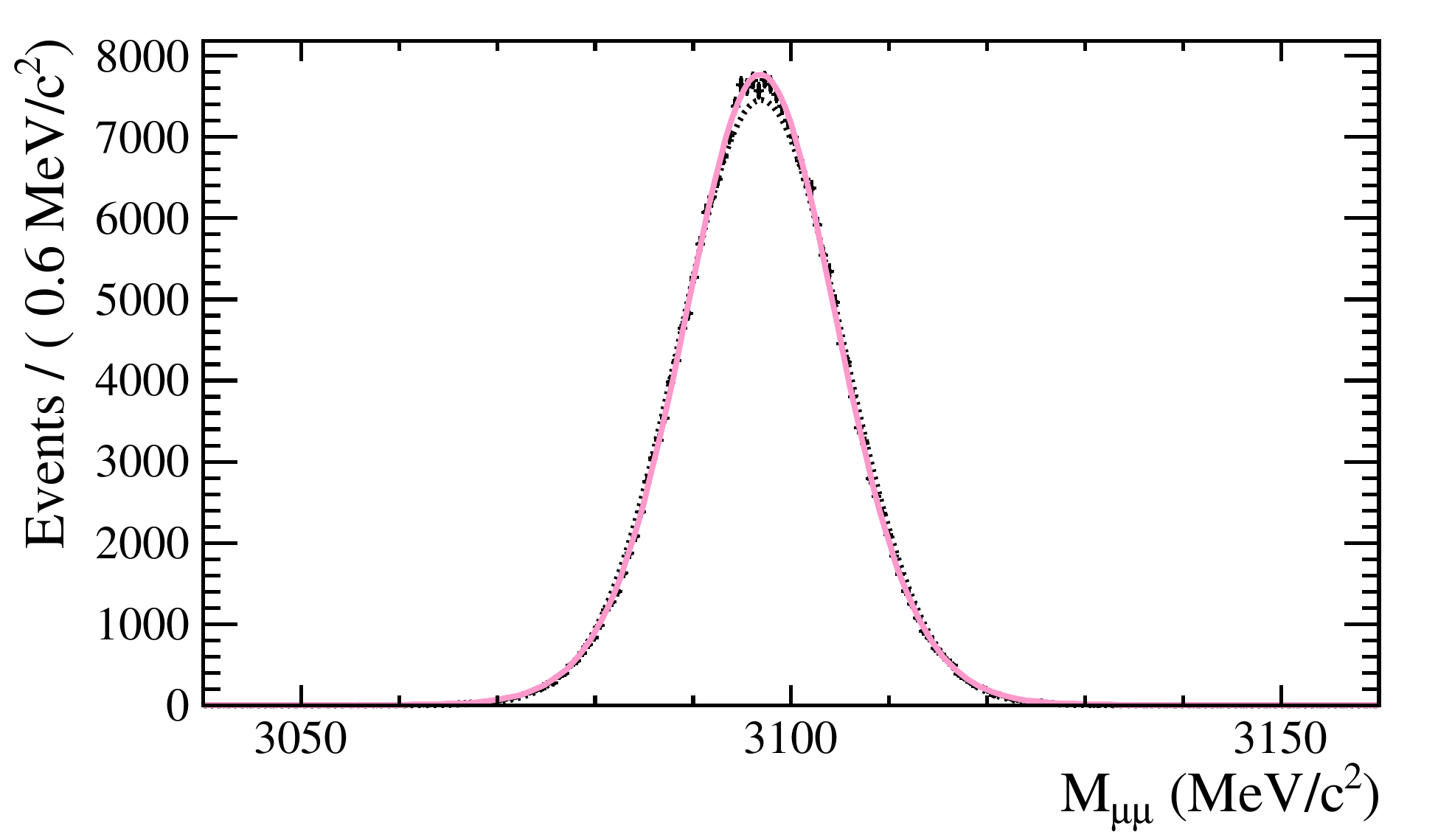}
\includegraphics[width=0.49\textwidth]{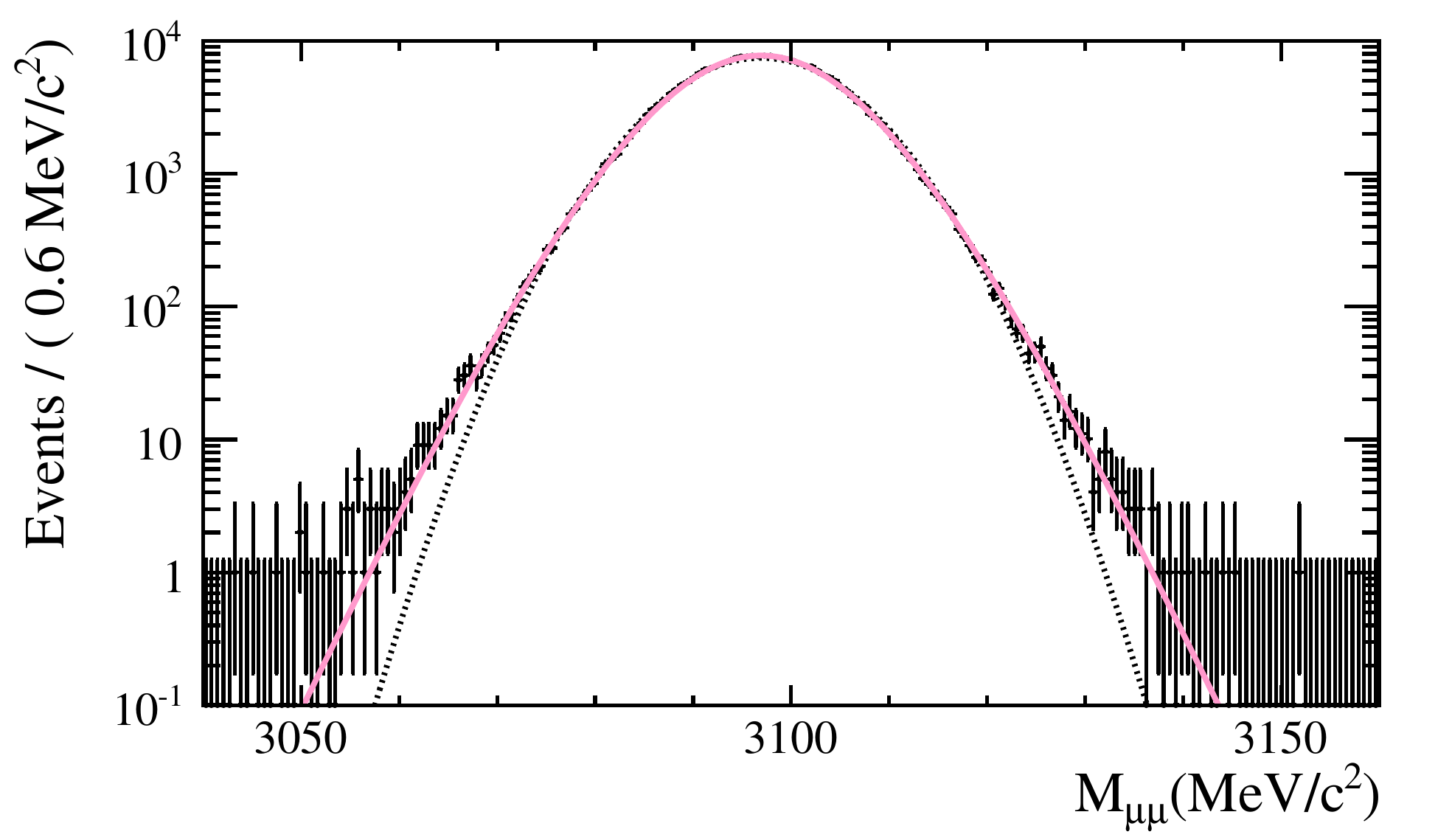}
\caption{Fit of the invariant mass distribution of a \Jpsimm generated sample with $p_{T}^{\Jpsi} \in [0,14]$\gevc, $\theta_{\mu} \in [20,300]$ mrad,
$p^{\mu} > 6$ \gevc and $p_{T}^{\mu} > 0.5$\gevc. The pink line corresponds to the fit to a hyperbolic distribution. The dashed black line corresponds to the
fit to a Gaussian. Left: linear scale. Right: Logarithmic scale.}
\end{center}
\end{figure}

Unfortunately, the above integral cannot be solved analytically.  It would require a numerical implementation of the core and its derivative. This would make the matching with the radiative tail difficult. Evaluating \eqref{eq:phi} numerically for different values of Amoroso parameters, we find
log-densities that exhibit an hyperbolic profile. Based on that observation, we define a possible core:

\begin{equation}
\label{eq:hyp_core}
c(x) \propto e^{-b\sqrt{1+(m-\mu)^2/\delta^2}}
\end{equation}

\noindent i.e, $c(x)$ is the symmetric limit of the hyperbolic distribution. It can also be rewritten
in such way that the mass resolution $\sigma$ appears explicitly, as it will be discussed in \secref{sec:ghyp}.
Adding a CB-like tail to \eqref{eq:hyp_core}, we obtain the following \pdf:
\begin{equation}
A(m,\mu,b,\delta,a,n) \propto
\begin{cases}
e^{-b\sqrt{1+(m-\mu)^2/\delta^2}} &\text{, if } \frac{m-\mu}{\delta} \geq -a\\
e^{-b\sqrt{1+a^2}}\left(\frac{n\sqrt{1+a^2}}{ba(n\sqrt{1+a^2}/(ba)-a-(m-\mu)/\delta)}\right)^n&\text{, otherwise}
\end{cases}
\end{equation}

\noindent hereafter referred to as the {\it Apollonios} distribution, currently being used for cross-checks
in data analysis of the LHCb experiment. The core \eqref{eq:hyp_core} can be obtained analytically for a variance prior density:
\begin{equation}
\label{eq:prior_hyp_core}
\rho(v,b,\delta) \propto e^{-(b^2v/\delta^2 + \delta^2/v)}
\end{equation}
We fit the mass peak for \Jpsimm decays satisfying  $p_{T}^{\Jpsi} \in [0,14]\gevc$, $\theta^{\mu} \in [20,300]$ mrad,
$p^{\mu} > 6\gevc$, $p_{T}^{\mu} > 0.5\gevc$ (which mimics LHCb-like conditions) to the {\it Apollonios} distribution,
and find a very good agreement as can be seen in \figref{fig:fit1}.

\begin{figure}
\label{fig:fit2}
\begin{center}
\includegraphics[width=0.49\textwidth]{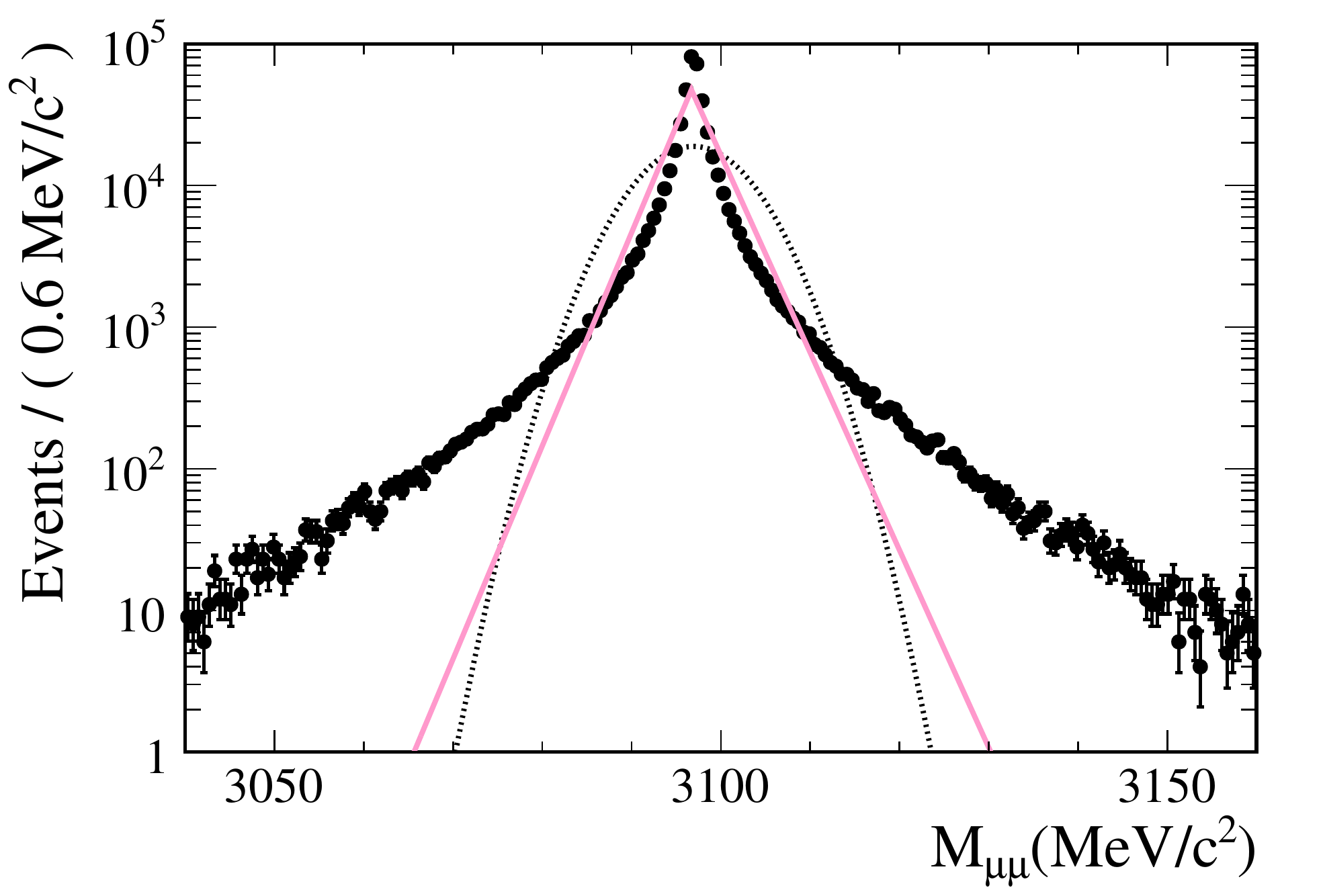}
\includegraphics[width=0.49\textwidth]{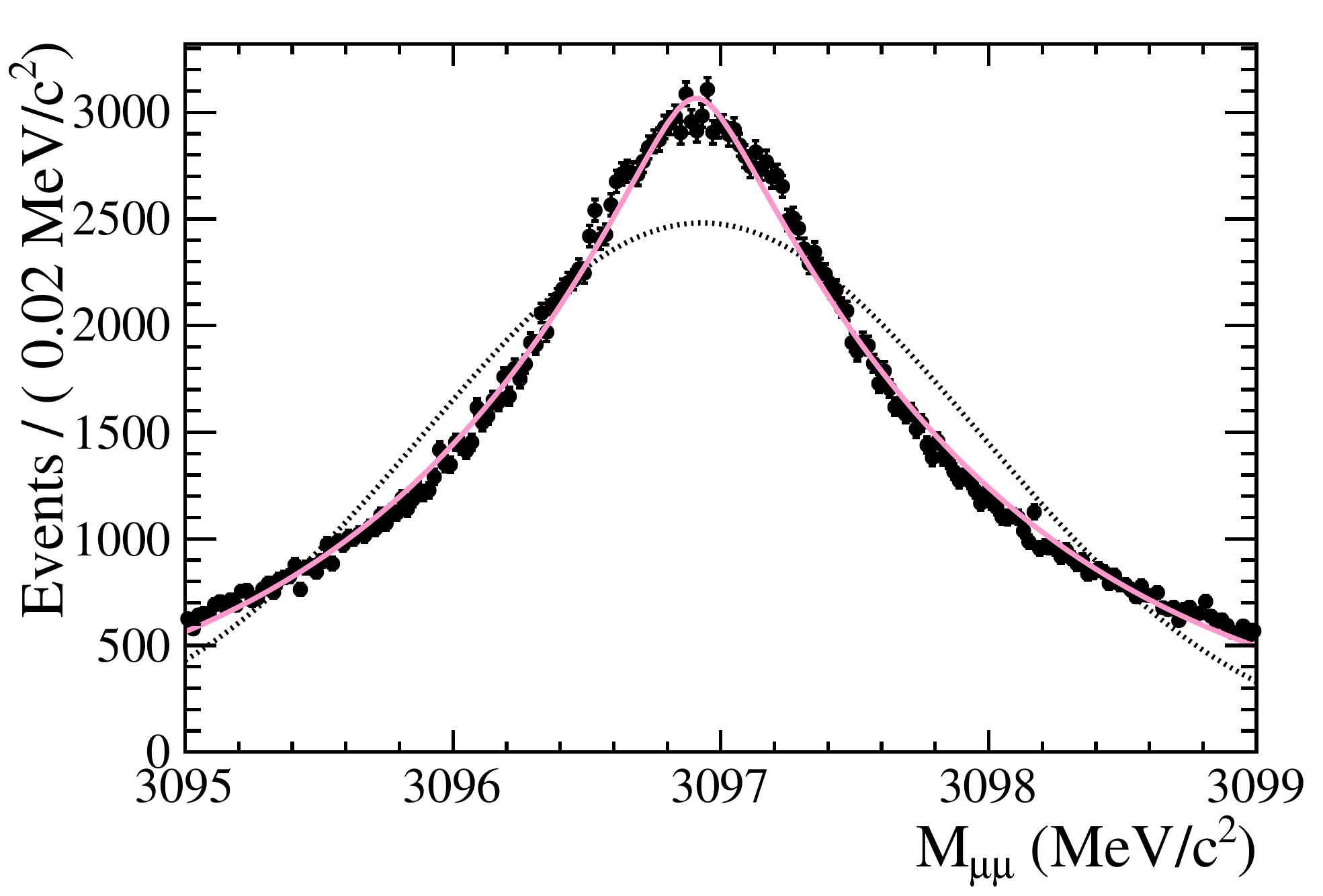}
\caption{Fit of the invariant mass distribution of a \Jpsimm generated sample without any phase space restriction and without the multiple
scattering term in the momentum resolution. The pink line corresponds to the fit to an hyperbolic distribution. The dashed black line corresponds to the
fit to a Gaussian. Left: Fit in the full mass range. Right: Fit in a region of about two standard deviations around the mean.}
\end{center}
\end{figure}

Now, the good agreement between this model and the MC toys used for testing can be broken without too much effort.
For example, we now repeat the exercise releasing all kinematic and acceptance cuts, and switching off the MS term.
These changes modify the distribution 
and fit results are shown in \figref{fig:fit2}, where we see that \eqref{eq:hyp_core} cannot fit the generated data. However,
it is also interesting to note that, even in this extreme case, \eqref{eq:hyp_core} can do a good job in a region of about two standard deviations
around the peak.

%% file: ghyp.tex
\section{Generalized Hyperbolic resolution model}
\label{sec:ghyp}

The core of~\eqref{eq:hyp_core} is a limit case of the generalized hyperbolic distribution~\cite{ghyp}:
\begin{equation}
\label{eq:ghyp}
G(m,\mu,\lambda,\alpha,\beta, \delta) = \left((m-\mu)^{2} + \delta^{2}\right)^{\frac{1}{2} \lambda - \frac{1}{4}} e^{\beta (m-\mu)} K_{\lambda - \frac{1}{2}}\left(\alpha \sqrt{(m-\mu)^{2} + \delta^{2}}\right)
\end{equation}
\noindent where $K_{\lambda}$ are the cylindrical harmonics or special Bessel functions of third kind. In principle, $\beta^2$ is constrained to be smaller than $\alpha^2$. In practice that
condition can be ignored if the fitting range is finite, but one has to be careful that if $\beta^2 > \alpha^2$ one of the tails will start rising at some point. The generalized 
hyperbolic distribution also has an important limit case, the {\it Student's-t} distribution, as indicated in \tabref{tab:GHYP}.

\begin{table}
\caption{Limit cases of the generalized hyperbolic distribution}
\label{tab:GHYP}
\begin{center}
\begin{tabular}{|c|c|}
 \hline
 Distributions & $G(m,\mu,\lambda,\alpha,\beta, \delta)$ \\
 \hline
 Hyperbolic & $\lambda=1, \alpha\delta = b$ \\
 Symmetric hyperbolic & $\lambda=1, \beta = 0, \alpha\delta = b$ \\
 Student's t & $\lambda=-\frac{\nu}{2} , \alpha=0, \beta = 0, \delta = \sqrt{\nu}$ \\
 Non-standardized Student's t & $\lambda=-\frac{\nu}{2} , \alpha=0, \beta = 0$ \\
 \hline
\end{tabular}
\end{center}
\end{table}

\begin{figure}
\label{fig:fit3}
\begin{center}
\includegraphics[width=0.49\textwidth]{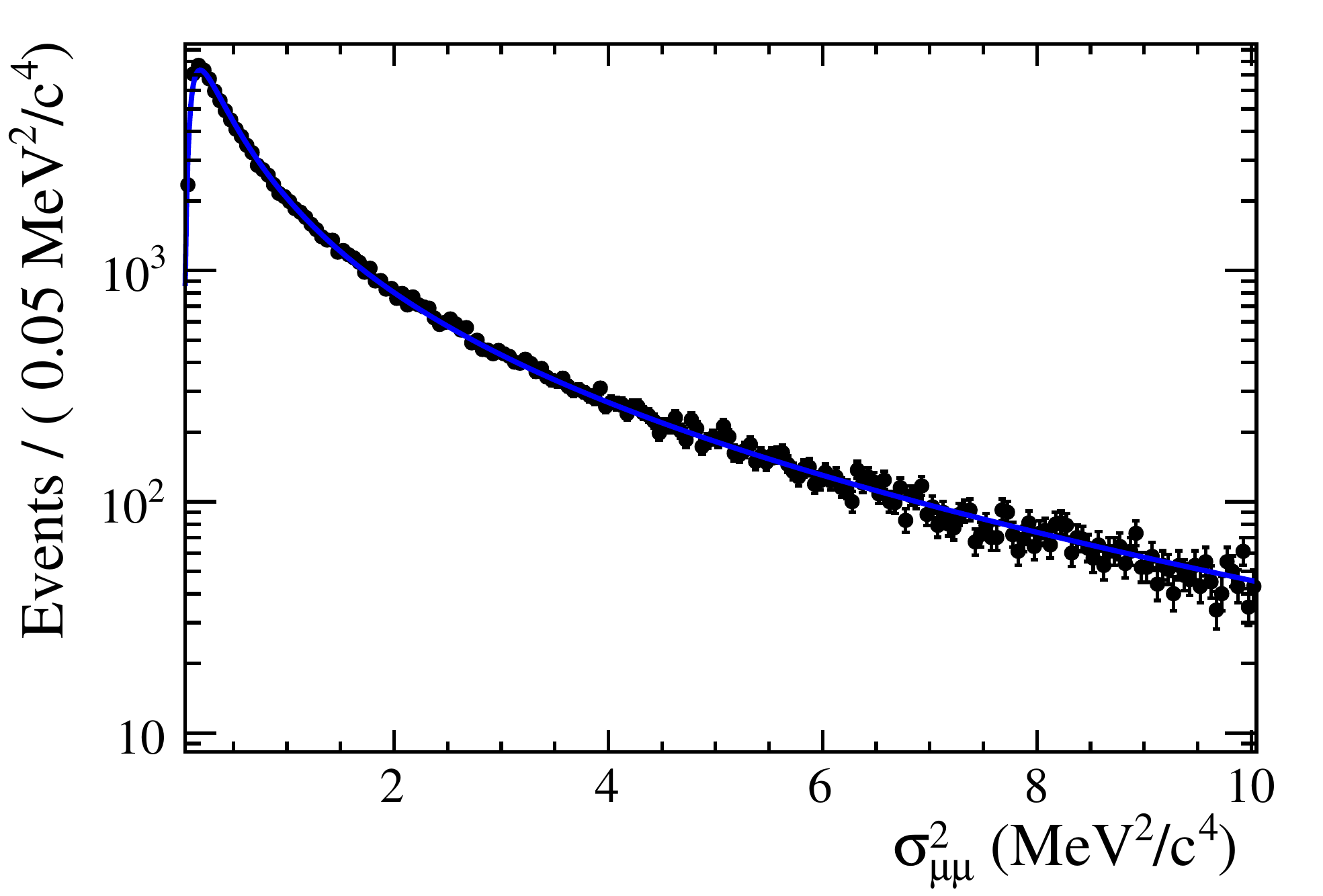}
\includegraphics[width=0.49\textwidth]{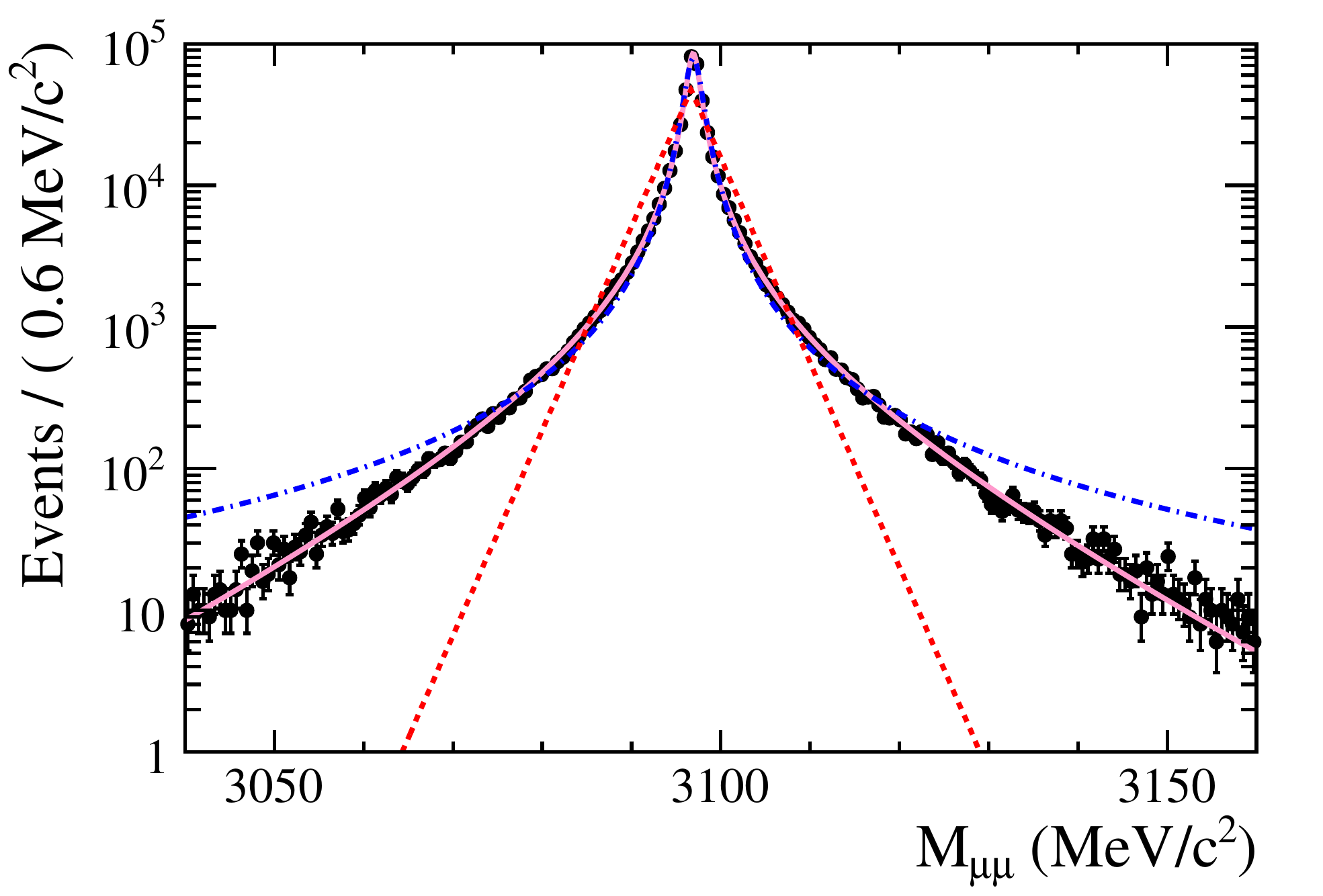}
\caption{Left: Per-event mass error squared fitted to ~\eqref{eq:rho_inv_gauss} in a \Jpsimm sample generated without MS.
Right: Fit to the mass distribution on the same sample. The pink solid line shows the generalized hyperbolic. The dot-dashed blue line the
{\it Student's-t } case, and the dashed red line the hyperbolic distribution. }
\end{center}
\end{figure}

The \pdf in \eqref{eq:ghyp} can also be obtained by marginalizing over a variance density\footnote{The parameter $\beta$ is related to a variance-dependency of the Gaussian mean, and not to the per event variance distribution. For the purpose of this paper $\beta$ can be considered zero.}:

\begin{equation}
\label{eq:rho_inv_gauss}
\rho(v,\lambda,\alpha,\delta) \propto v^{\lambda-1}e^{-[\alpha^2v + \delta^2/v]}
\end{equation}
The distribution \eqref{eq:rho_inv_gauss} is the generalized inverse Gaussian distribution and describes very well
the density we find for $\sigma_{\mu\mu}^2$, for the example in \figref{fig:fit2}. This is shown in
\figref{fig:fit3}, together with the good agreement between the simulated data and the generalized hyperbolic
distribution. We find that \eqref{eq:rho_inv_gauss} fits well the mass variance distribution for all the generated \Jpsimm\ samples that we have tested, although one needs to add an overall offset to the per-event error, i.e, to change $v$ by $v-v_0$ in
\eqref{eq:rho_inv_gauss}. The effect of an overall displacement of the per-event error distribution is further discussed in \secref{sec:offset}.

The following re-parametrization $\{\alpha;\delta\}\to\{\sigma;\zeta\}$:
\begin{align}
\zeta &= \alpha\delta \\
\sigma^2 &= \delta^2 \frac{K_{\lambda+1}(\zeta)}{\zeta K_{\lambda}(\zeta)} = \delta^2 A^{-2}_{\lambda}(\zeta)
\end{align}
\noindent is more suitable for fitting purposes as it allows us to specify the rms ($\sigma$) of the
distribution in the symmetric case ($\beta =0$) as an explicit parameter. $A^{2}_{\lambda} = \frac{\zeta K_{\lambda}(\zeta)}{K_{\lambda+1}(\zeta)}$ is introduced for further convenience.
In that parametrization:

\begin{multline}
\label{eq:ghypreparam}
G(m,\mu,\sigma,\lambda,\zeta,\beta) \propto \\
\left((m-\mu)^{2} + A^{2}_{\lambda}(\zeta)\sigma^{2}\right)^{\frac{1}{2} \lambda - \frac{1}{4}} e^{\beta (m-\mu)} K_{\lambda - \frac{1}{2}}\left(\zeta \sqrt{1+(\frac{m-\mu}{A_{\lambda}(\zeta)\sigma})^{2}}\right)
\end{multline}

\begin{figure}
\label{fig:ghyp}
\begin{center}
\includegraphics[width=1.\textwidth]{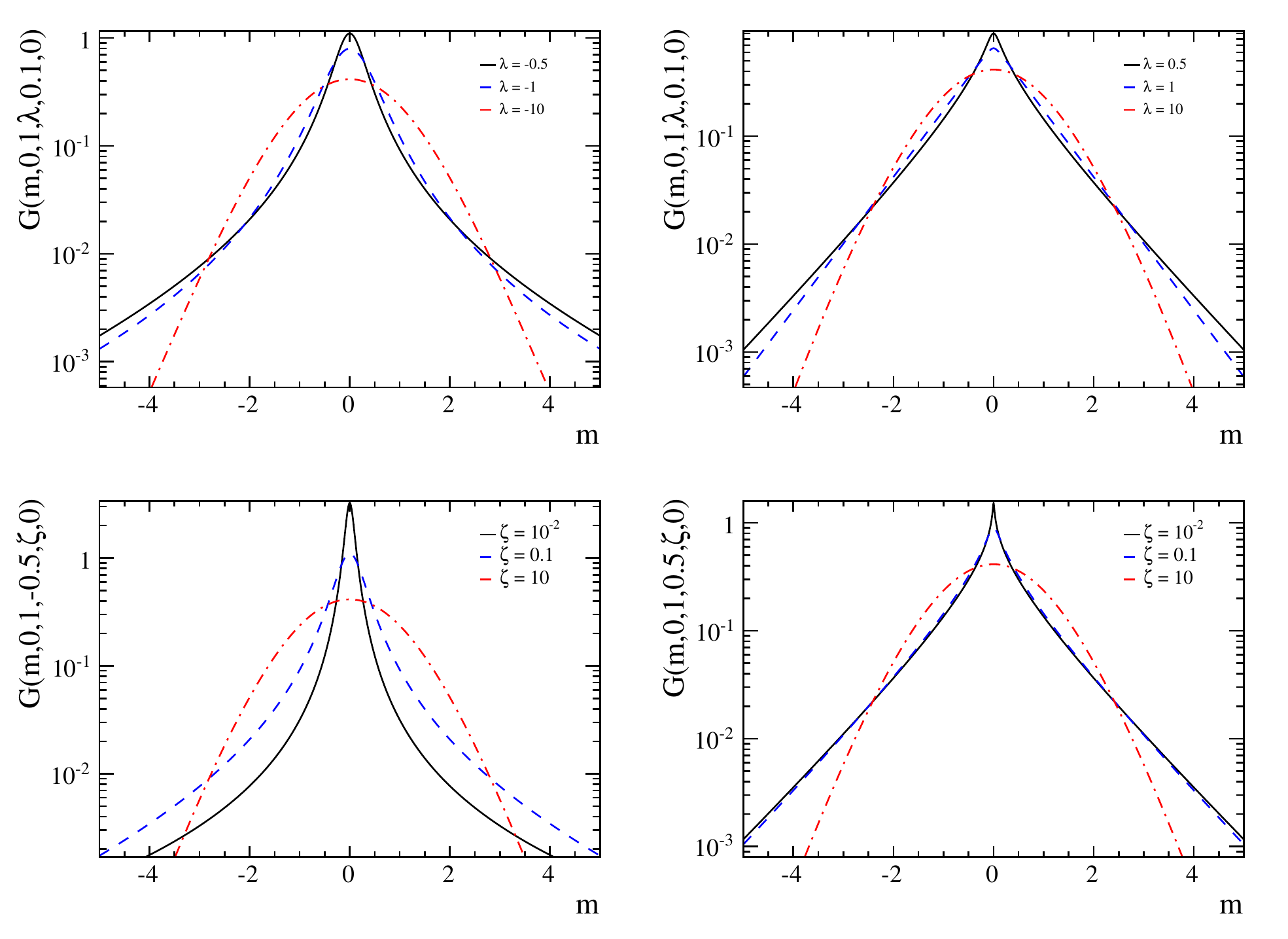}
\caption{$G(m,\mu,\sigma,\lambda,\zeta,\beta)$ is plotted for standard values of $\mu$, $\sigma$, $\beta$ and different values of $\zeta$ and $\lambda$. \textit{Top}: $\zeta$ is fixed to 0.1 and $\lambda$ is varied. \textit{Bottom}: $\lambda$ is fixed to -0.5 for the \textit{left} plot and 0.5 for the \textit{right} plot and $\zeta$ is varied.}
\end{center}
\end{figure}

Figure \ref{fig:ghyp} shows $G(m,\mu,\sigma,\lambda,\zeta,\beta)$ for different values of $\zeta$ and $\lambda$.

Using \eqref{eq:ghypreparam} as the core of a CB-like function, we define:
\begin{multline}
I(m,\mu,\sigma,\lambda,\zeta,\beta,a,n) \propto \\
\begin{cases}
\left((m-\mu)^{2} + A^{2}_{\lambda}(\zeta)\sigma^{2}\right)^{\frac{1}{2} \lambda - \frac{1}{4}} e^{\beta (m-\mu)} K_{\lambda - \frac{1}{2}}\left(\zeta \sqrt{1+(\frac{m-\mu}{A_{\lambda}(\zeta)\sigma})^{2}}\right) &\text{, if } \frac{m-\mu}{\sigma} > -a\\
\frac{G(\mu-a\sigma,\mu,\sigma,\lambda,\zeta,\beta)}{\left(1-m/(n\frac{G(\mu-a\sigma,\mu,\sigma,\lambda,\zeta,\beta)}{G'(\mu-a\sigma,\mu,\sigma,\lambda,\zeta,\beta)}-a\sigma)\right)^n} &\text{, otherwise}
\end{cases}
\end{multline}

\noindent hereafter referred to as {\it Hypatia} distribution, where $G'$ is the derivative of the $G$ defined in \eqref{eq:ghyp}.

The generalized hyperbolic core can describe most of the examples that were generated (see \figref{fig:fit3}, right),
but can also be broken with high statistic samples if \Jpsimm events are taken all over the phase
space without any kinematic or acceptance requirement, as it can be seen in \figref{fig:Ypsilon}.

%% file: offset.tex
\section{Effect of the offset}
\label{sec:offset}
We have seen that \eqref{eq:rho_inv_gauss} is a flexible function that can parametrize
mass variance distributions if an offset is added to it. Yet, by adding the offset, the marginalization
does not yield a generalized hyperbolic distribution for the most general case.
We can see that adding an offset to the per-event error distribution is equivalent to performing a convolution:
\begin{multline}
\label{eq:convolution}
\int_0^{\infty} \frac{1}{\sqrt{v}}e^{-\frac{1}{2v}(m-\mu)^2} \rho(v-v_0) dv = \int_0^{\infty} \frac{1}{\sqrt{v_0+\Delta}}e^{-\frac{1}{2(v_0+\Delta)}(m-\mu)^2} \rho(\Delta) d\Delta = \\
= \int_0^{\infty} \left(\int_{-\infty}^{+\infty}\frac{1}{\sqrt{v_0}}e^{-\frac{1}{2(v_0)}(m-t)^2}\frac{1}{\sqrt{\Delta}}e^{-\frac{1}{2(\Delta)}(t-\mu)^2} dt \right) \rho(\Delta) d\Delta = \\
=  \int_{-\infty}^{+\infty}\frac{1}{\sqrt{v_0}}e^{-\frac{1}{2(v_0)}(m-t)^2}\left(\int_0^{\infty}\frac{1}{\sqrt{\Delta}}e^{-\frac{1}{2(\Delta)}(t-\mu)^2} \rho(\Delta) d\Delta \right) dt = \\
= \frac{1}{\sqrt{v_0}}e^{-\frac{1}{2(v_0)}(m)^2}\ast\left(\int_0^{\infty}\frac{1}{\sqrt{\Delta}}e^{-\frac{1}{2(\Delta)}(t-\mu)^2} \rho(\Delta) d\Delta \right) 
\end{multline}

\begin{figure}
\label{fig:Ypsilon}
\begin{center}
\includegraphics[width=0.9\textwidth]{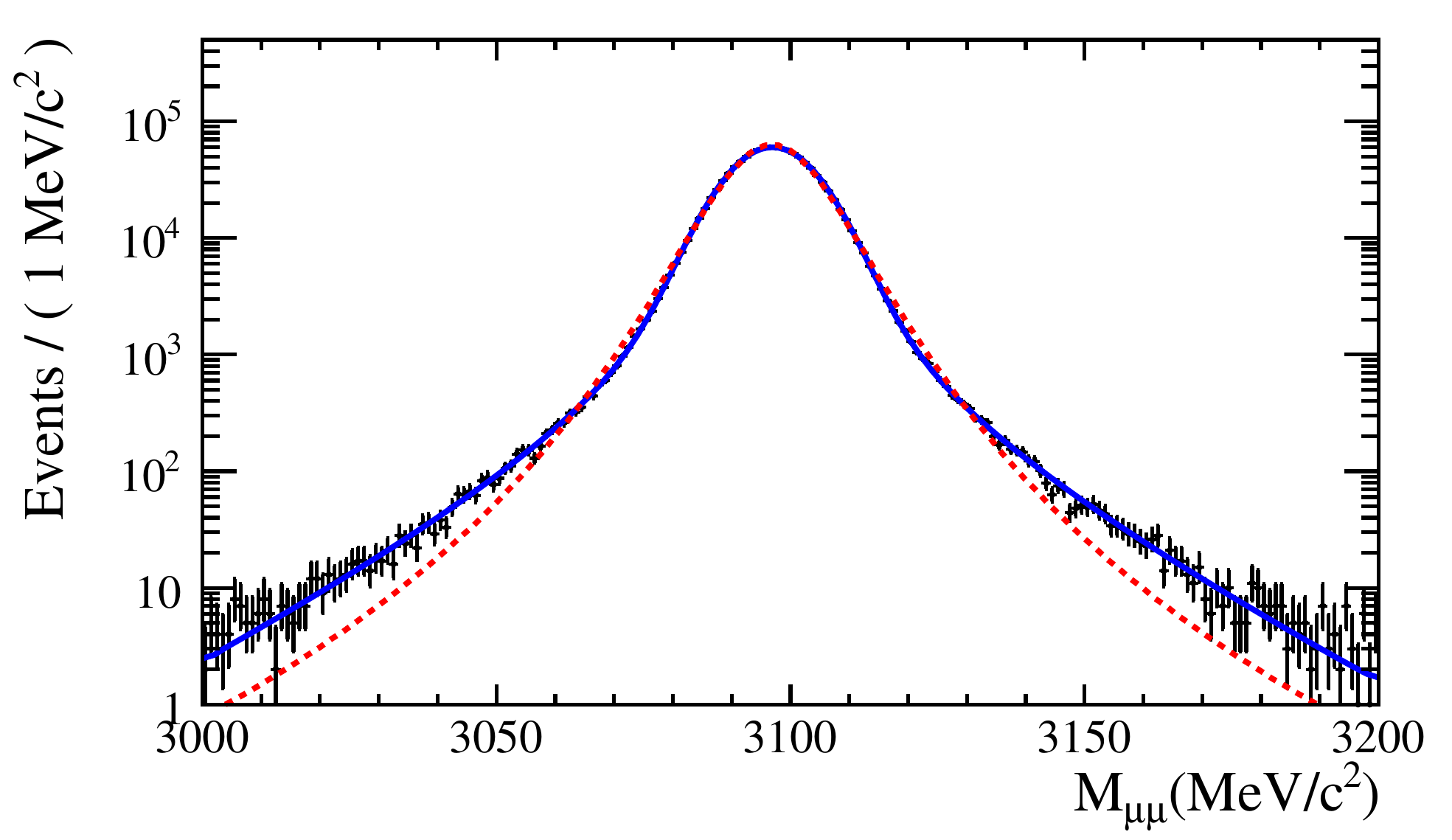}
\caption{Fit of $I(m,\mu,\sigma,\lambda,\zeta,0,\infty,1)$ (red dashed) and $\Upsilon(m,\mu,\sigma^{SR},\lambda,\zeta,0,\infty,1,6.5\MeVcc)$ (solid blue) to the simulated \Jpsimm data, for the full sample without any
kinematic constraint.}
\end{center}
\end{figure}

The convolution of a generalized hyperbolic distribution with a Gaussian is not in general another generalized hyperbolic. However, we can argue that if $v_0>>\Delta$, we will
have a single Gaussian (that is a limit case of the generalized hyperbolic) and, on the contrary, that if $v_0<< \Delta $ in most of the $\Delta$ range, we will recover the generalized hyperbolic distribution.
One can also argue that as we are looking for corrections to the Gaussian distribution, the convolution properties of the Gaussian function still hold approximately.
Yet, it will not be exact, and therefore a smeared {\it Hypatia} distribution:
\begin{equation}
\Ypsilon(m,\mu,\sigma^{SR},\lambda,\zeta,\beta,a,n, v_0) =  \frac{1}{\sqrt{v_0}}e^{-\frac{1}{2(v_0)}(m)^2}\ast I(m,\mu,\sigma^{SR},\lambda,\zeta,\beta,a,n)
\end{equation}
\noindent can provide a better fit than $I(m,\mu,\sigma,\lambda,\zeta,\beta,a,n)$ for some complicated cases with high statistics, without a real increase in 
the number of fit parameters ($\sqrt{v_0}$ can be fixed in a somewhat arbitrary point at the start-up of the mass error distribution), although
at the cost of a numerical convolution. The later can be done in \RooFit by calling the  \texttt{RooFFTConvPdf} class on top of the implementation
of $I(m,\mu,\sigma,\lambda,\zeta,\beta,a,n)$. If written this way, $\sqrt{v_0}$ can be interpreted as an estimate of the mass resolution due to multiple scattering, $\sigma^{SR}$ the dispersion
caused by the spatial resolution of the detector given the kinematics of the final state, and the total resolution would be $\sigma = \sqrt{v_0 + (\sigma^{SR})^2}$.
\figref{fig:Ypsilon} shows a fit of $I(m,\mu,\sigma,\lambda,\zeta,0,\infty,1)$ and $\Upsilon(m,\mu,\sigma^{SR},\lambda,\zeta,0,\infty,1,6.5\MeVcc)$ to the simulated \Jpsimm data, for the full sample without any
kinematic constraint, i.e., where very low momentum (MS dominated) and very high momentum (hit resolution dominated) coexist. The fitting
range corresponds to about eleven standard deviations. 
An excellent agreement between $\Upsilon(m,\mu,\sigma^{SR},\lambda,\zeta,0,\infty,1,6.5\MeVcc)$  and the simulated data is found. 
We failed to find any subsample of the \Jpsimm data that could not be fitted by $\Ypsilon(m,\mu, \sigma^{SR},\lambda,\zeta,0,\infty,1, v_0)$. 

%% file: ranges.tex
\section{Properties of $\lambda$}
\label{sec:ranges}

We have seen that using the Hypatia $\Ypsilon$ distribution we can factorize the mass resolution modelling into MS and SR. The first part is governed
by a resolution parameter $\sigma_0$ that can be estimated from the start-up of the per-event variance distribution. The second part is governed
by the parameters $\zeta,\lambda,\sigma$ of the generalized hyperbolic distribution, where $\sigma$ corresponds to the resolution introduced by SR 
and where, empirically, we have found that $\zeta$ is in most cases small. In this chapter we will derive a physical meaning for $\lambda$, at least
in the small $\zeta$ limit. 

\begin{figure}
\label{fig:sigma_vs_p}
\begin{center}
\includegraphics[width=0.45\textwidth]{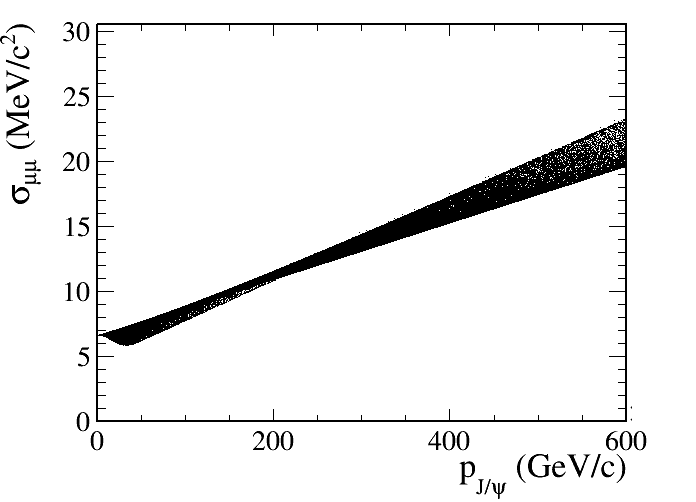}
\includegraphics[width=0.45\textwidth]{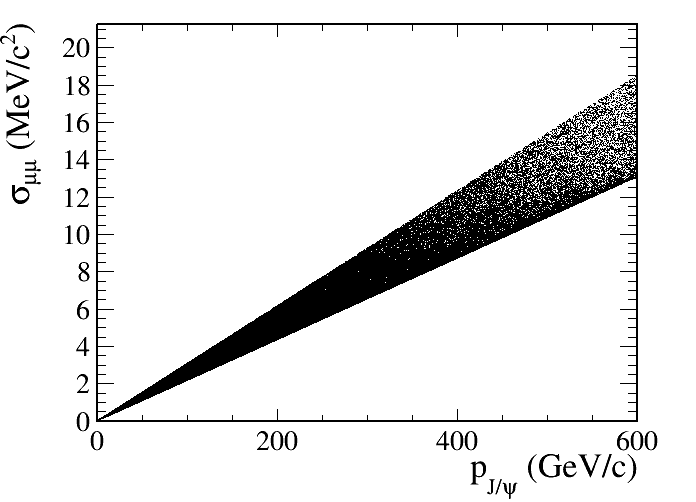}
\caption{Per-event mass uncertainty versus \Jpsi momentum. Left: with multiple scattering. Right: only detector resolution.}
\end{center}
\end{figure}

In the $\alpha = 0 $ ($\rightarrow\zeta = 0$) limit case, the generalized hyperbolic distribution becomes a {\it Student's-t} distribution, which can
be understood as a marginalization over a per-event variance density:
\begin{equation}
\label{eq:inv_gamma}
\rho(v) \propto v^{\lambda-1}e^{-b/v}
\end{equation}
The mean ($M$) and mode ($\mu$) of \eqref{eq:inv_gamma} are:

\begin{equation}
M = \frac{b}{-\lambda-1}; \mu = \frac{b}{-\lambda+1}
\end{equation}

\noindent thus
\begin{equation}
\lambda = \frac{1+M(v)/\mu(v)}{1-M(v)/\mu(v)} < 0
\end{equation}
\noindent and we can get an estimate of $\lambda$ by looking at the per-event error (squared) distribution,
and making the ratio of its mean and mode after shifting it to start at zero. But, we can further exploit
this relation. From \eqref{eq:smearing} we can suppose that the per-event uncertainty will be
strongly correlated with the particle momenta. Indeed, \figref{fig:sigma_vs_p} supports this.

\begin{equation}
\sigma_i^{SR} \approx c_{te} \times p_{\Jpsi,i}
\end{equation}

If this is the case, then $M(v^{SR})/\mu(v^{SR}) \approx M(p_{\Jpsi}^2)/\mu(p_{\Jpsi}^2)$ and $\lambda$ does not
depend on detector effects, only on particle kinematics. This is an interesting result, because if we have a MC
simulation with a good description of the momentum distribution of the particles in the lab frame, then the 
values of $\lambda$ obtained in simulation should be reasonably valid for data, regardless of having an accurate description of detector simulation.

\begin{table}
\caption{Results of a fit to $\Ypsilon(m,\mu,\sigma^{SR},\lambda,0,0,\infty,1,\sigma^{MS})$ for toy MC $\Jpsi$ events smeared
with different values of $a$ and $b$ in \eqref{eq:smearing}. The parameter $\sigma^{MS}$ is fixed at the start-up of the per-event variance.
The parameter $\lambda$ is found to be very stable with respect to smearing parameters, which are varied by $100\%$. 
However the uncertainty on $\lambda$ varies significantly, and increases with $a/b$.}
\begin{center}

\begin{tabular}{|c|c||c|c|c|}
 \hline
a$[\times10^{-3}]$ & b $\gevc^{-1}$ & $\lambda$ &$\sigma^{MS} (\mevcc)$&$\sigma^{SR} (\mevcc)$\\
 \hline
3 & 2$\times10^{-5}$ & $-2.40\pm0.06$ & 6.81 & $4.75\pm0.02$ \\
1.5 & 2$\times10^{-5}$ & $-2.10\pm0.03$ & 3.53 & $3.71\pm0.01$ \\
6 & 2$\times10^{-5}$ & $-2.67\pm0.16$ & 13.3 & $6.3\pm0.05$ \\
3 & 4$\times10^{-5}$ & $-2.11\pm0.03$ & 7.07 & $7.53\pm0.03$ \\
3 & 1$\times10^{-5}$ & $-2.65\pm0.15$ & 6.67 & $3.06\pm0.03$ \\
 \hline
\end{tabular}
\end{center}
\end{table}

%% file: constraint.tex
\section{Mass constraints on intermediate resonances}
\label{sec:const}

Up to now we have described resolution effects. In more complicated cases,
it is sometimes very useful to apply constraints
on the decay products. For example, one can significantly improve the invariant mass resolution of \BsJpsiPhi by constraining the two muons
to have the PDG \Jpsi mass \cite{PDG}. This kind of approach, although great at 
improving the overall resolution, can also generate tails on the mass 
distribution, due to the photon energy radiated in the \Jpsimm decay.
Let's consider a simple case in which the constraint is 
just applied by substituting the mass of the dimuon by the mass of the
\Jpsi.
\begin{equation}
m_{c}^2 = m_{\Jpsi}^2 + m_{KK}^2 + 2 \left(\sqrt{m_{\Jpsi}^2 + p_{\mu\mu}^2} \sqrt{m_{KK}^2 + p_{KK}^2} -p_{\mu\mu}p_{KK}\cos(\theta)\right)
\end{equation}
while ideally one would have wanted to implement:
\begin{equation}
m_{true}^2 = m_{\Jpsi}^2 + m_{KK}^2 + 2 \left(\sqrt{m_{\Jpsi}^2 + p_{\Jpsi}^2} \sqrt{m_{KK}^2 + p_{KK}^2} -p_{\Jpsi}p_{KK}\cos(\theta)\right)
\end{equation}

The difference $m_c^2-m_{true}^2$ is not zero but rather function of the energy of the photons generated in the \Jpsimm decay. This difference 
can be greater than zero, generating a tail on the right-hand side. 
Hence, even with a perfect detector resolution, the combination of the mass constraint and the photon
radiation will generate non-Gaussian tails. In practice, this effect is expected to be small because the \Jpsimm decays are selected
with a mass window cut that allows only low energy photons. Otherwise, it can be partially accommodated either by the
resolution model (e.g \eqref{eq:ghyp}) or by using a CB-like tail on the right-hand side (i.e, using a double-sided {\it Hypatia}).
A further discussion that goes beyond the scope of this paper is to provide models marginalized
over per-event errors.

%% file: conclusions.tex
\section{Conclusions}
\label{sec:conclusions}

We have presented a generalization of the Crystal Ball function that gives an excellent description
of mass resolution non-Gaussian tails. This function, that we name the {\it Hypatia} distribution, $I$,  corresponds to a CB-like tail with a generalized
hyperbolic core. The smeared {\it Hypatia} distribution, $\Ypsilon$ provides an improved description of mass peaks and its
fit parameters have clearer fundamental meaning, although the price to pay is a numeric convolution.
A second, right-hand side CB-like tail can be added in cases where one has other non-resolution effects,
such as those coming from constraining the mass of intermediate resonances of a decay. 

%% file: ack.tex
\section{Acknowledgments}
\label{sec:ack}

We would like to thank L. Carson and V. Gligorov for his useful comments
on this draft. We would also like to thank W. Hulsbergen for helpful
discussions during the preparation of this work.